\documentclass[conference]{IEEEtran}
	\IEEEoverridecommandlockouts
	\usepackage{cite}
	\usepackage{multirow}
	\usepackage{amsmath,amssymb,amsfonts}
	\usepackage{mathrsfs}
	\usepackage{algorithmic}
	\usepackage{graphicx}
	\usepackage{subfig}
	\usepackage{textcomp}
	\usepackage{xcolor}
	\usepackage[font=small]{caption}
	\def\BibTeX{{\rm B\kern-.05em{\sc i\kern-.025em b}\kern-.08em
			T\kern-.1667em\lower.7ex\hbox{E}\kern-.125emX}}
	\begin{document}
		
		\title{Compressive Sensing based Multi-class Privacy-preserving Cloud Computing 
		}
		
		\author{\IEEEauthorblockN{Gajraj Kuldeep, Qi Zhang}
			\IEEEauthorblockA{\textit{DIGIT, Department of Engineering, Aarhus University, Denmark} \\
				Email:\{gkuldeep, qz\}@eng.au.dk\\
				\\
			\textit{Accepted in IEEE Global Communications Conference 2020}}
		}
		
		\maketitle
		
		\begin{abstract}
			In this paper, we design the multi-class privacy$\text{-}$preserving cloud computing scheme (MPCC) leveraging compressive sensing for compact sensor data representation and secrecy for data encryption. The proposed scheme achieves two-class secrecy, one for  superuser who can retrieve the
			exact sensor data, and the other for semi-authorized user who is only able to obtain the statistical data such as mean, variance, etc. MPCC scheme allows computationally expensive sparse signal recovery to be performed at cloud without
			compromising the confidentiality of data to the cloud service providers. In this way, it mitigates the issues in data transmission, energy and storage caused by massive IoT sensor data as well as the increasing concerns about IoT data privacy in cloud computing. Compared with the state-of-the-art schemes, we show that MPCC scheme not only has lower computational
			complexity at the IoT sensor device and data consumer, but also is proved to be secure against ciphertext-only attack.

		\end{abstract}
		
		\begin{IEEEkeywords}
			privacy preserving, compressed sensing,  encryption, cloud computing, and
			IoT 
		\end{IEEEkeywords}
		
		\section{Introduction}
		
		The internet of things (IoT) ecosystem enables massive sensor data acquisition, transmission, and storage, as well as computation, to perform data analytic for diverse smart monitoring, process automation, and control services. The ever-increasing data generation imposes many critical challenges to communication, storage, and computing infrastructure\cite{IoTsurvey}. Cloud can be used for storage and computing; however, there are increasing concerns about data privacy in cloud \cite{DSPPI}. Data privacy issues can be handled using homomorphic encryption schemes, but it requires computation-intensive operations at the encoder\cite{homosurvey}. Therefore, it is challenging to use homomorphic encryption schemes in resource-constrained IoT sensors.  
		
		Compressive sensing (CS) is a promising solution to mitigate the above-mentioned problems in the IoT ecosystem. CS has laid the foundation for the recovery of sparse and compressible signals using fewer measurements, thereby providing efficient sampling and compact representation\cite{bib:EJT,bib:DLD}. CS has been used for simultaneous compression and encryption ~\cite{cse1,cse2}.
		CS has been used to provide multi-class encryption schemes and also allows the cloud to perform computationally expensive sparse signal recovery.
		
\begin{figure}[htbp]
			\centering
			\includegraphics[width=.9\linewidth]{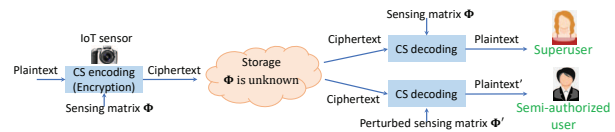}
			\caption{ Multi-class encryption architecture with cloud storage for multi-class data consumers.}
			\label{multi}
		\end{figure}

		Multi-class encryption using CS has been demonstrated in~\cite{multi1,multi2,multi3}. Service model for multi-class encryption scheme is shown in Fig. \ref{multi}. 
		 The authors\cite{multi2,multi1} introduced multi-class users by perturbing the entries of the binomial sensing matrix at the encoding processing. The superuser knows complete sensing matrix, therefore perfect signal reconstruction at the superuser.
The semi-authorized user has the perturbed sensing matrix. Therefore, the reconstructed signal at the semi-authorized user is of low quality or without sensitive information. Synchronization of sensing matrices for multi-class users and at the IoT device is required for signal recovery, which results in cumbersome key management from IoT device to multi-class users. Multi-class encryption for images is proposed in~\cite{multi3}, where the perturbation key is watermarked into the sensed image, which simplifies the key management for the multi-class encryption scheme. Note that the sensing matrices are not known to the cloud; therefore, the schemes in \cite{multi1,multi2, multi3} cannot perform the computationally expensive decompression at the cloud.

\begin{figure}[htbp]
		 	
		 	\includegraphics[width=1\linewidth]{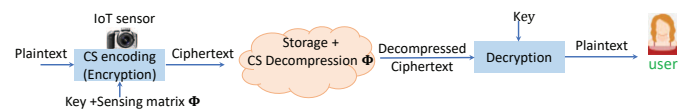}

		 	\caption{ Service model for cloud assisted storage and decompression.}	\label{cld}
		 \end{figure}

		CS encoding is a linear process, whereas reconstruction of the signal from the measurements is nonlinear and computationally expensive. It is preferable to use the cloud for both storage and signal reconstruction. Service model for cloud-assisted storage and decompression is shown in Fig. \ref{cld}.
		 Since  signal reconstruction at the cloud compromises data privacy, additional privacy-preserving mechanism should be added in the encoding process.
		 
		In the literature, CS-based cloud computing of image data for efficient storage and reconstruction while maintaining the privacy was proposed in\cite{PAOIRS}, and the cloud is assumed to be semi-trusted. Cloud-assisted health care monitoring  using CS was proposed in\cite{PACAHM}.  Both the schemes \cite{PAOIRS,PACAHM} provided decompression of encrypted measurements at the cloud. However, to apply the decompression algorithm at the cloud, the obfuscated sensing matrix needs to be transmitted from the data consumer to the cloud, thereby adding transmission cost for the data consumer. Furthermore, the size of the stored data is more than the original data because secrecy is provided by adding noise to the CS measurements, which makes these schemes inefficient for storage.
		Cloud assisted CS-based data gathering was proposed in ~\cite{PPDC,SCSDGS}. The encoding process can be described in the following way to achieve confidentiality at the cloud~\cite{PPDC}. The signal is premultiplied with a random permutation matrix then on the permuted signal CS-based sensing is applied. Afterwords, the CS measurements are  multiplied  by a random matrix. Again in this scheme for decompression at the cloud the data consumer needs to send the obfuscated sensing matrix to the cloud.
		In ~\cite{SCSDGS}, the confidentiality of data at the cloud is achieved by incorporating the multiplication of a permutation matrix before sensing and corrupting a few entries of measurement vector. In this scheme, sensing matrix is known to the cloud and the decompressed signal is stored at the cloud. 
		
		Our recent study shows that CS-based encryption schemes for binomial and Gaussian sensing matrices are not secure against ciphertext-only attack in resource-constrained IoT sensor devices\cite{OurP}. In this paper, we applied the ciphertext-only attack\cite{OurP} on the schemes in\cite{multi2,multi3,SCSDGS} and observed that these scheme can leak information.   We propose Multi-class Privacy-preserving Cloud Computing scheme (MPCC) that fulfills multi-class encryption as well as privacy-preserving cloud computing. {It can have diverse applications, e.g., reversible face de-identification in the video footage. The semi-authorized user receives video without sensitive blocks whereas the superuser receives the original video.}  
		
		Our contributions can be summarized as follows. i) We propose MPCC scheme which is integrated with the CS framework to provide compression and privacy-preserving cloud storage. ii) MPCC scheme allows the data decompression at the cloud without compromising the data secrecy. iii) MPCC scheme is compared with the state-of-the-art schemes in terms of computation and security. iv) Theoretical security analysis of  MPCC scheme is performed. It  is also proved that it is computationally infeasible to break the proposed scheme. v) Empirical results are demonstrated for different data types. 
		
		The paper is organized as follows: Section II contains the basics of CS. Section III and IV describe service and threat model and  MPCC  scheme, respectively. Section V includes  security analysis of MPCC scheme. Section VI compares  MPCC scheme with the existing work. Applicability of  MPCC scheme on smart meter data and image is shown in Section VII. Finally, Section VIII concludes the paper. 
		\\
		\textit{Notations:} In this paper, all boldface uppercase and all boldface lowercase  letters represent matrices and vectors, respectively. $\mathbf{x}^T$ is transpose of $\mathbf{x}$.
		Italic letters represent variables. Calligraphic letters represent set. 
		\section{Preliminaries}
		In this section, the basics of CS are presented. Let $\mathbf{x}$ be a signal which is either exactly  $K$-sparse in the canonical form or approximately sparse in the transform domain. An exactly $K$-sparse signal is defined as $||\mathbf{x}||_0=K$ whereas an approximately sparse signal, $\mathbf{x=\Psi \theta}$, is defined as $||\mathbf{\theta}||_0=K$ which means most of signal information is contained in the $K$  coefficients of   $\mathbf{x}$'s transformation.
		We represent $l_p$ norm of a vector $\mathbf{x}$ as $(\sum_{i=1}^{N}|x_i|^p)^{\frac{1}{p}}$. 
		
		Compression is achieved by taking random linear measurements using a sensing matrix.  CS measurements are given as,
		\begin{eqnarray}
		\mathbf{y=\Phi x}, \label{csb}
		\end{eqnarray}
		where $\mathbf{\Phi}\in \mathbf{R}^{M\times N}$ is a sensing matrix.
		It has been shown that the sensed signal can be recovered using convex optimization if the signal satisfies the sparsity constraint, and the sensing matrix satisfies the restricted isometric property (RIP)~\cite{bib:rip}. 
		%
		%
		
		If the entries of the sensing matrix are chosen from i.i.d. Gaussian distribution and the number of measurements should be in the order of $Klog(N/K)$~\cite{ec2}, then it satisfies the RIP with probability one. If the matrix satisfies the RIP then the signal can be recovered using $l_1$ minimization by solving the following optimization problem,
		\begin{eqnarray}
		\mathbf{\hat{x}}=\arg \min \limits_{\mathbf{x} \in \mathbb{C}^{N}} \mathbf{||x}||_1, \text{       s.t.   } \mathbf{y}=\mathbf{\Phi} \mathbf{x},  
		\end{eqnarray} 
		where the signal $\mathbf{x}$ is sparse in canonical form~\cite{bib:EJT,bib:DLD}. In the case when signal is approximately sparse, then the optimization problem becomes,
		\begin{eqnarray}
		\mathbf{\hat{\theta}}=\arg \min \limits_{\mathbf{\theta} \in \mathbb{C}^{N}} \mathbf{||\theta}||_1, \text{       s.t.   } \mathbf{y}=\mathbf{\Phi\Psi} \mathbf{\theta}.  
		\end{eqnarray}
		In the next section we present  the CS-based service model and threat model for multi-class encryption and cloud computing.
		
		\section{Service model and Threat model}
\begin{figure}[htbp]
			\centering
			\includegraphics[width=3.4in]{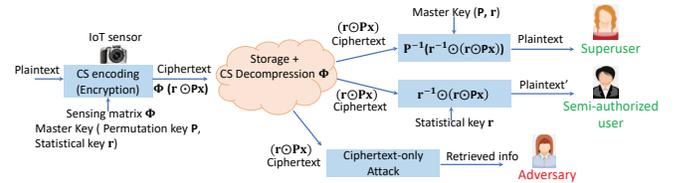}
			\caption{The proposed architecture for multi-class encryption and outsourcing of computationally expensive end-user decoding to cloud.}
			\label{newDia}	
		\end{figure}

		The service model that allows cloud to perform computationally expensive CS decompression, while providing multi-class encryption is shown in Fig.~\ref{newDia}.
	
		The IoT device continuously senses data using CS, and the sensed data is transmitted to the cloud and stored at the cloud. We authorize two
		types of user consumers, i.e., superuser and semi-authorized user, to have different data access right using different keys. The superuser has the master key and can retrieve plaintext, whereas the semi-authorized user can only retrieve permuted
		plaintext with the statistical key. 
		To retrieve data, the data consumer  sends a query to the cloud, then the cloud performs sparse signal recovery and sends the decompressed ciphertext to the data consumer. The data consumers can apply their key on the  ciphertext and recover the corresponding signal information according to their access right. 
		Data privacy is maintained while data is transmitted  and stored at the cloud using the proposed scheme. 
		
		For the multi-class encryption and computation outsourcing, privacy challenges need to be addressed.  
		Secrecy challenges can be divided into  end-to-end secrecy, privilege escalation, and storage secrecy categories.
		
		End-to-end secrecy: Data transmission secrecy should be achievable from the IoT device to the superuser or semi-authorized user. Privilege escalation: To guarantee that the semi-authorized user will not be able to recover the information, which is only meant for the superuser.
		Storage secrecy: Since sensor data is stored at the cloud, and decompression is also performed by the cloud. It should prevent the cloud from recovering part of the data. In the proposed scheme, we consider the cloud as a potential adversary who follows the communication protocol but could have an intention to retrieve plaintext information.
		\section{Multi-class privacy-preserving cloud computing scheme}
		In this section, we design the Multi-class Privacy-preserving Cloud Computing scheme (MPCC) that addresses the threat scenario in Section III while providing
		multi-class encryption and computation outsourcing to the cloud.  Public parameters for the MPCC are $N$, $M$, $K$ and $\mathbf{\Phi}$, where $N$ is signal length, $M$ is the number of measurements, $K$ is the sparsity of the signal, and $\mathbf{\Phi}$ is the sensing matrix. MPCC is represented as MPCC=\{\textit{Keygen, EncE, CCom, DecSU, DecSAU}\}. Description of the  MPCC is given in the following.
		
		\textit{Keygen} is the key generation algorithm that generates key $K_r$ and $K_p$. These keys are derived using  master secret. 
		
		\textit{EncE} is the encryption algorithm at the IoT device that generates permutation matrix $\mathbf{P}_i$ and random vector $\mathbf{r}_i$ using  $K_p$ and $K_r$, respectively.  Permutation matrix and random vector can be generated using stream cipher for keys $K_p$ and $K_r$, respectively. Using the $\mathbf{P}_i$ and $\mathbf{r}_i$, ciphertext of message $\mathbf{x}_i$, $\mathbf{y}_i$ is given as,
		\begin{equation}
		\mathbf{y}_i=\mathbf{\Phi (r}_i\odot\mathbf{P}_i \mathbf{x}_i),\label{encE}
		\end{equation}
		where $\odot$ is Hadamard product. Ciphertext $\mathbf{y}_i$ and the corresponding index is sent to cloud.
		
	    \textit{CCom} is the sparse recovery algorithm at the cloud which decompresses the ciphertext. Cloud performs decompression on demand either from the superuser or the semi-authorized user. Cloud decompresses the ciphertext using the public parameters and request query, the decompressed signal  for the query of $i^{th}$ data block is given as,
		\begin{equation}
		\mathbf{z}_i=\mathbf{r}_i\odot\mathbf{P}_i \mathbf{x}_i.
		\end{equation}   
		\textit{DecSU} is the decryption algorithm at the superuser. It takes key $K_r$, $K_p$ and the decompressed signal, $\mathbf{z}_i$, as input to reconstruct the signal. The reconstructed signal at the superuser is given as,
		\begin{equation}
		\mathbf{\hat{x}}_i=\mathbf{P}_i^{-1}(\mathbf{r}_i^{-1}\odot(\mathbf{r}_i\odot \mathbf{P}_i \mathbf{x}_i)).\label{decSU}
		\end{equation}
		\textit{DecSAU} is the decryption algorithm at the semi-authorized user, only has key $K_r$ and the decompressed signal, $\mathbf{z}_i$, as input to reconstruct the signal. The reconstructed signal at the semi-authorized user is given as,
		\begin{equation}
		\mathbf{x}^{p}_i=\mathbf{r}_i^{-1}\odot(\mathbf{r}_i\odot \mathbf{P}_i\mathbf{x}_i).\label{decSAU}
		\end{equation}
		The reconstructed signal at the semi-authorized user is the permutation of the original signal. Hence, the semi-authorized
		user is unable to know the plaintext but able to calculate the statistical values of the plaintext.
		
		\section{Security analysis of the MPCC scheme} 
		In this section, we show that the MPCC scheme is computationally secure. We also show that computationally secrecy of the  MPCC scheme can be increased by modifying the public parameters without changing the base scheme. It should be noted that secrecy of the MPCC scheme is not achieved through sensing matrix in CS. CS is used for compression and sampling.  It can be observed from encryption Eq. \ref{encE}, for the cloud or other adversary, the security of the proposed scheme depends on the permutation of the plaintext and the Hadamard product of random stream with plaintext. To show that these operations are sufficient to provide computational secrecy, we demonstrate security analysis as follows. 
		
		Let $\mathbb{F}$ be a finite field with additive and multiplicative identities as $\mathscr{I}_a$ and $\mathscr{I}_m$, respectively. The plaintext space, the ciphertext space, and   the key space are given as  $\mathbb{M=F}^N$,  $\mathbb{C=F}^M$, and  $\mathbb{K}=\mathbb{F}^N-\{\mathscr{I}_a\}$, respectively.
		For plaintext $\mathbf{x}\in \mathbb{M}$, ciphertext is given as  
		\begin{eqnarray}
		\mathbf{c=r\odot Px=r\odot x'}, \label{tte}
		\end{eqnarray}
		where $\mathbf{r,P}\in \mathbb{K}$ and $\mathbf{x}'$ is permutation of the plaintext $\mathbf{x}$. It can be observed that ciphertext given in Eq. \ref{tte} is simplified version of Eq. \ref{encE}. Considering only elements of $\mathbf{c}$, which are not equal to $\mathscr{I}_a$, we can design encryption as, Enc: $c=rx'$, where $x'$, $r$, and $c$ are element of $\mathbf{x'}$, $\mathbf{r}$, and $\mathbf{c}$, respectively. The decryption is designed as, Dec: $r^{-1}c=x'$. The probability for a given $x'$ and $r$ is chosen uniformly, we have,
		\begin{eqnarray}
		\Pr\limits_{r\in \mathbb{K}}(Enc(x',r)=c)&=&\frac{\text{ \# Keys such that } Enc(x',r)=c}{\text{Total number of keys}},\nonumber\\
		&=&\frac{1}{|\mathbb{F}|-1},
		\end{eqnarray}
		where $|\mathbb{F}|$ is the number of elements in the field.
		Let $x'_0$ and $x'_1$ be two elements of the plaintext, then we have,
		\begin{eqnarray}
		\Pr\limits_{r\in \mathbb{K}}(Enc(x'_0,r)=c)=\Pr\limits_{r\in \mathbb{K}}(Enc(x'_1,r)=c).
		\end{eqnarray}
		This shows the elements of $\mathbf{c}$, which are not equal to $\mathscr{I}_a$, are perfectly secure\cite{pf}. As this can also be observed from the Table \ref{fieldT} that any ciphertext $c$ occurs with probability  $\Pr(c)=\frac{1}{4}$ for field  $\mathbb{F}_5$. Indices of plaintext are secured by the permutation matrix. Hence by incorporating Hadamard product and permutation at the encoder, we achieve perfect secrecy for the values of plaintext and computational secrecy for the indices of plaintext.
		\begin{table*}
			\subfloat[]{\centering
				\begin{tabular}{|l|l|l|l|l|}
					\hline
					r\textbackslash{}m & 1 & 2 & 3 & 4 \\ \hline
					1 & 1 & 2 & 3 & 4 \\ \hline
					2 & 2 & 4 & 1 & 3 \\ \hline
					3 & 3 & 1 & 4 & 2 \\ \hline
					4 & 4 & 3 & 2 & 1 \\ \hline
				\end{tabular}
				\label{fieldT}}
			\subfloat[]
			{
				\centering{
					\begin{tabular}{|c|c|c|c|c|c|}
						\hline
						& \cite{PACAHM} &\cite{SCSDGS}  & \cite{multi2} &\cite{multi3}& \begin{tabular}[c]{@{}l@{}}MPCC\end{tabular} \\ \hline
						\begin{tabular}[c]{@{}l@{}}Cloud storage  cost (in Bytes)\end{tabular} &$ O(2N)$ & $O(N)$ & $O(M)$ & $O(M)$ & $O(M)$   \\ \hline
						\begin{tabular}[c]{@{}l@{}}Sensor data Transmission\end{tabular} & $O(M)$ & $O(M)$ & $O(M)$ & $O(M)$ & $O(M)$\\ \hline
						Sensor computation & $O(MN)$ & $O(MN)+[\Psi]$ & $O(MN)$& $O(MN)$ & $O(MN)+[\Psi]$ \\ \hline
						\begin{tabular}[c]{@{}c@{}}Data transmission for\\ query from superuser\end{tabular} & $O(MN)$ & Simple query & N/A & N/A & Simple query \\ \hline
						\begin{tabular}[c]{@{}c@{}}Superuser  computation\\ (in operation)\end{tabular} & \begin{tabular}[c]{@{}l@{}}$O(N^{\theta})$\\  $2<\theta <3$ \end{tabular} & $O(N)+[\Psi]$ & $O(N^3)$ & $O(N^3)$ & $O(N)+[\Psi]$ \\ \hline
						Privacy protection & \begin{tabular}[c]{@{}l@{}}Permuted  index of data\end{tabular} & \begin{tabular}[c]{@{}l@{}}Permuted and\\  changed  value\end{tabular} & CS &
						\begin{tabular}[c]{@{}c@{}} CS+\\Watermarking \end{tabular} & \begin{tabular}[c]{@{}l@{}}Permuted and\\ changed  value\end{tabular} \\ \hline
						Semi-authorized user & N/A & N/A & $O(N^3)$ & $O(N^3)$ & $O(N)+[\Psi]$\\ \hline
				\end{tabular}}
				\label{comT}
			}
			\caption{(a) Ciphertext values for field $\mathbb{F}_5$. (b) Comparison of the MPCC scheme with  state-of-the-art schemes. N/A means that scheme does not support that functionality. }
		\end{table*}
		Now we calculate the computational cost to retrieve plaintext  at the semi-authorized user. From Eq. \ref{decSAU}, we know that at the semi-authorized user the plaintext values are known but not the indices. The number of operations required to apply brute force attack  for retrieving the plaintext indices with signal sparsity $K$ is given as, 
		\begin{eqnarray}
		\frac{N!}{N-K!}&=&N(N-1)(N-2)\dots(N-K+1),\nonumber \\
		&>& (N-K+1)^K.
		\end{eqnarray}
		Computational cost for the semi-authorized user is $O(N^K)$ to retrieve the indices of the plaintext from the ciphertext.

		Most of the natural signals take values from the real field. Let ${r_i\in}\{\pm a_1,\pm a_2,\dots,\pm a_T\}$ and $a_i$'s are chosen uniformly. Ciphertext at the cloud is, 
		$\mathbf{c=r\odot Px}$.
		Therefore to remove the effect of $\mathbf{r}$  from $\mathbf{c}$, cloud needs to perform $(2T)^K$ computation. At the cloud, the total number of computations to apply brute force attack is $O((2T)^KN^K)$ which is computationally infeasible to perform even for small values. For example, for $N=256$, $K=30$ and $T=16$ cloud needs to perform  $O(2^{390})$ computations. Despite that for positive signals the complexity is  $O(T^KN^K)$, by selecting permutation matrix's entries positive and negative with equal probability  then again computational complexity to apply brute force attack at the cloud is  $O((2T)^KN^K)$ even for positive signals.
		
		\section{Computational complexity}
		In this section, computational complexity for the MPCC scheme is evaluated at the IoT device, the superuser and the semi-authorized user. We use the nomenclature from  \cite{SCSDGS} to calculate computational complexity of the MPCC scheme. Computational complexity at the cloud is $O(N^3)$ because of $l_1$ minimization. Computation cost at the IoT device is $O(MN)$  as it can be observed from Eq. \ref{encE} for exactly $K$-sparse signals and it is $O(MN)+[\Psi]$ for signals which are sparse in transform domain. Signal at the superuser and the semi-authorized user is received in the decompressed form and to retrieve plaintext Hadamard product is performed as can be observed from Eq. \ref{decSU} and Eq. \ref{decSAU}, which requires $O(N)$ computations. The inverse of the permutation matrix at the superuser does not increase the complexity, keeping the complexity as $O(N)$. Therefore, overall computation remains  $O(N)$ for the superuser and the semi-authorized user. However, if signal is sparse in some transform domain then the computation cost is  $O(N)+[\Psi]$ for both superuser and semi-authorized user. 
		
		We also compare the MPCC scheme with the schemes presented in \cite{multi2}, \cite{multi3}, \cite{PACAHM}, and \cite{SCSDGS}. The comparison results are given in the Table \ref{comT}. The schemes in \cite{multi2,multi3} are designed using CS for multi-class users where the computationally expensive sparse signal recovery is performed at the data consumers. Whereas in the MPCC both storage and decompression is performed at the cloud, thereby reducing the requirement of resources at the data consumers. It requires transmission of an entire matrix from the user consumer to the cloud in\cite{PACAHM}, furthermore, it does not support multi-class user, whereas MPCC requires simple query to let cloud identify which block to decompress. The scheme in \cite{SCSDGS} has equivalent computation cost of MPCC; however,  it does not support multi-class users as MPCC does.

		\section{Experimentation}
		In this section, the MPCC scheme is applied to smart meter data and image. The proposed scheme is simulated using MATLAB and CS $l_1$-MAGIC optimization toolbox. 
		
		\subsection{Smart meter data}
		In this subsection, we consider the scenario of power meter reading at the access point from apartments  and access point transmitting these readings in the encrypted and compressed form to the cloud. At the cloud these compressed and encrypted readings are stored. These readings can be accessed by two types of users: superuser with the master key can reconstruct exact meter readings and semi-authorized user can reconstruct only the permuted readings thereby preserving the privacy of the customers. 
		\begin{figure}[htpb]
			\centering
			\includegraphics [width=.7\linewidth]{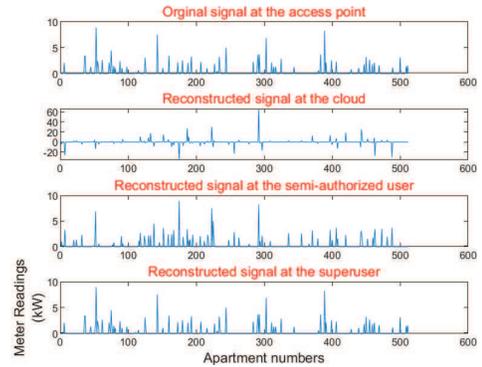}
			\caption{Smart power meter readings at the access point  and reconstructed signals at the cloud, superuser, and  semi-authorized user.}
			\label{powerR}
		\end{figure}
		
		For simulation, exactly sparse signals are constructed using apartment power usage from the smart meter dataset\cite{databasepower}. We choose first 70  apartments, readings and signal length 512 to construct the sparse signals. These readings are updated in every 15 minutes. Superuser's key is $(K_r,K_p)$ and semi-authorized user's key is $K_r$. We use Eq. \ref{encE} for compression and encryption at the IoT device, and $l_1$ minimization for decompression of encrypted data at the cloud. 
		\begin{figure}[htbp]
			\centering
			\subfloat[]{	\includegraphics [width=.49\linewidth]{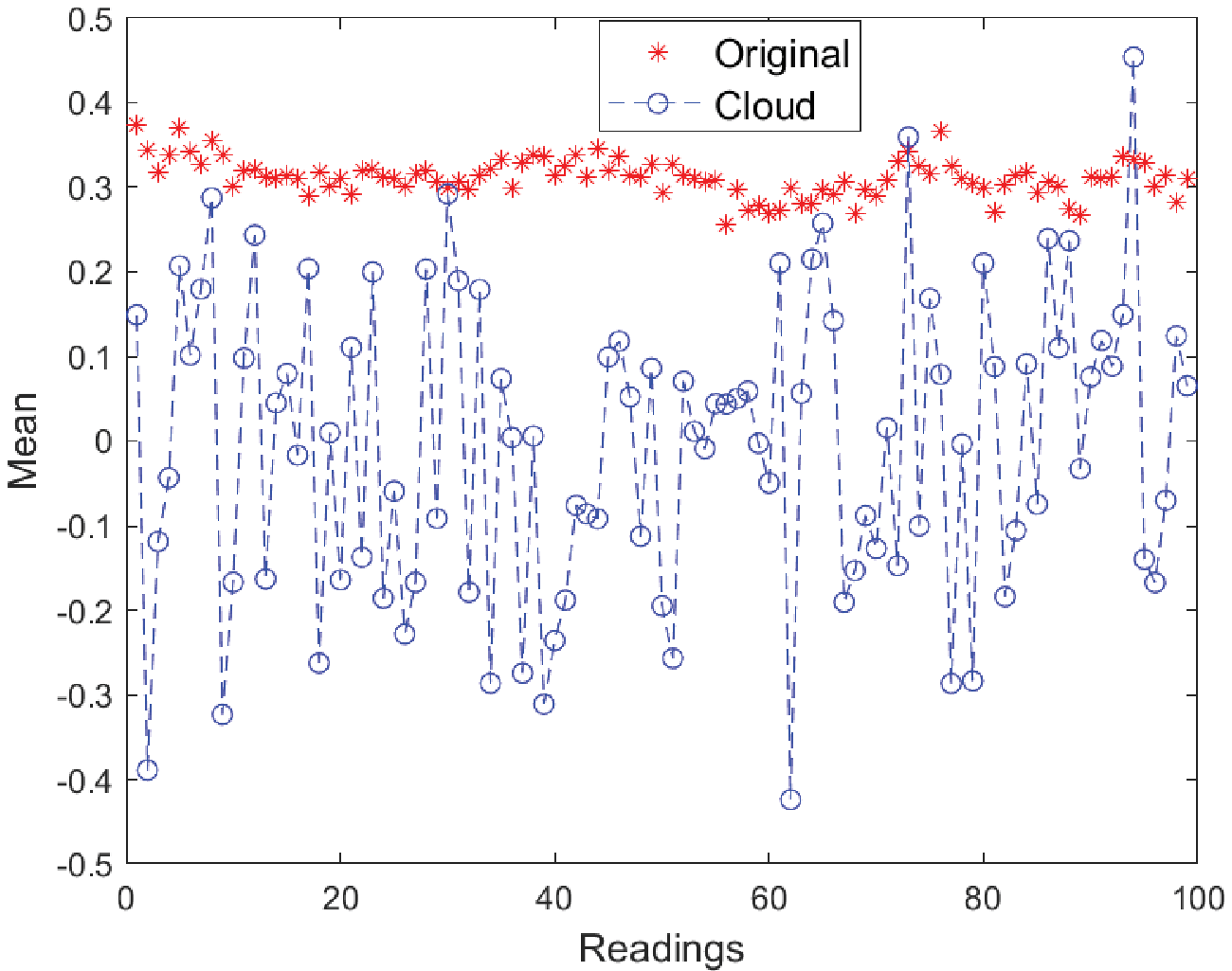}
				
				\label{SC}}
			\subfloat[]{\includegraphics [width=.49\linewidth]{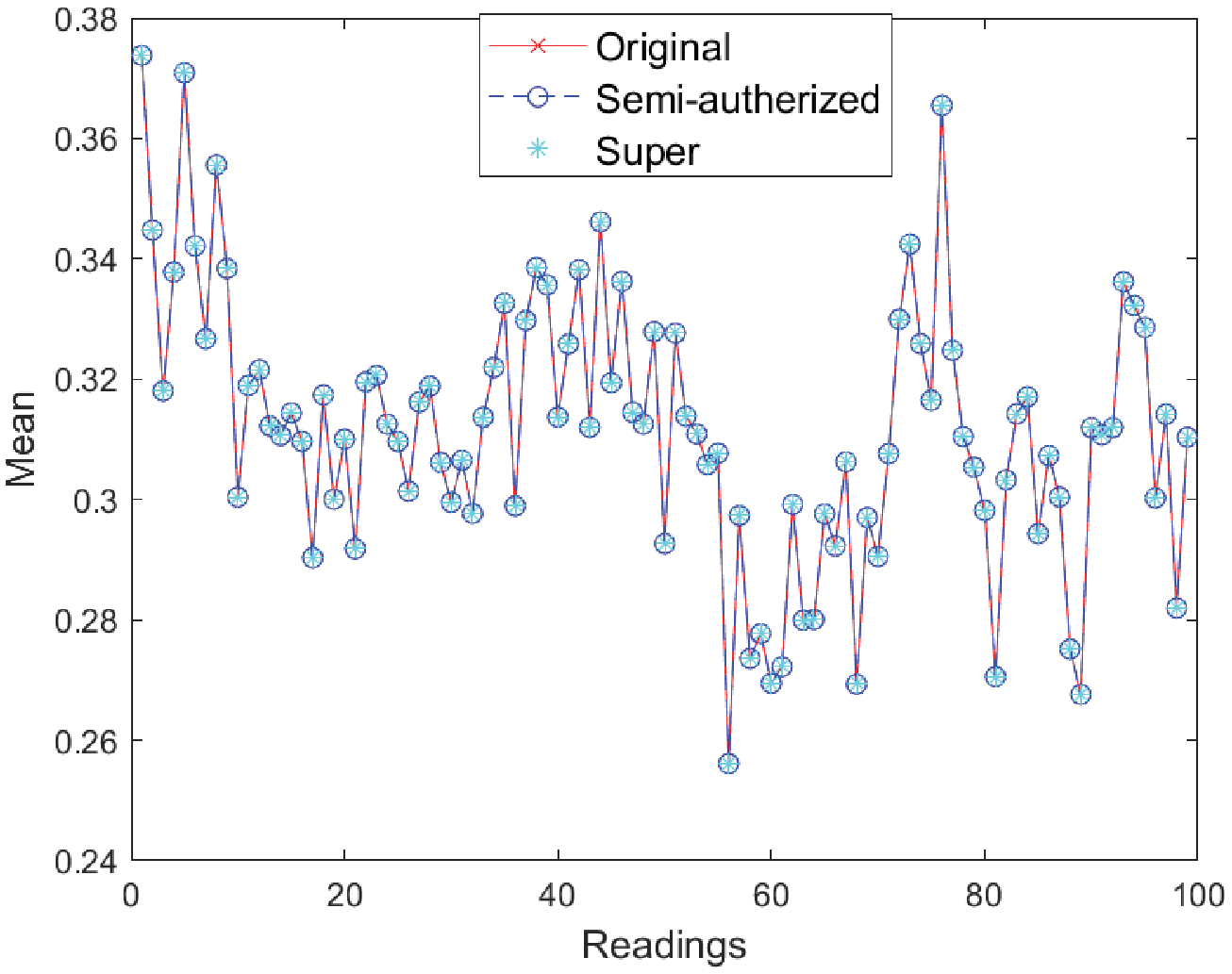}
				
				\label{mean2}}
			\caption{(a) Mean of the original signal and  reconstructed signal at the cloud. (b) Mean of the original signal and  reconstructed signal at the superuser and semi-authorized user.}	
		\end{figure}
		Fig. \ref{powerR} shows the original meter readings at the access point, as well as the reconstructed readings at the cloud, semi-authorized user, and superuser. 
		Semi-authorized user reconstructs the permuted meter readings with original value; hence, the relation between apartment and meter reading cannot be determined.
		Nevertheless, semi-authorized user can calculate statistical values on the permuted data such as, mean, variance, histogram etc, i.e., obtaining information of a group of IoT devices. In Fig. \ref{SC} the mean value of the original and reconstructed signal at the cloud is shown. From this figure, it can be observed that the  mean values calculated at the cloud are completely different from the original values, i.e., cloud is not able to retrieve any useful information. In Fig. \ref{mean2} the mean values at the semi-authorized and superuser are shown. It shows that the semi-authorized user like superuser can calculate the statistical values of power signal without any error.
		\subsection{Image}
		In this subsection, we  present security analysis of the schemes in \cite{SCSDGS,multi2,multi3}  on image signals. We show that the MPCC scheme can also be used for images to provide multi-class encryption. 
		
\begin{figure}[htbp]
			\centering
			\subfloat[]{	\includegraphics [width=.3\linewidth]{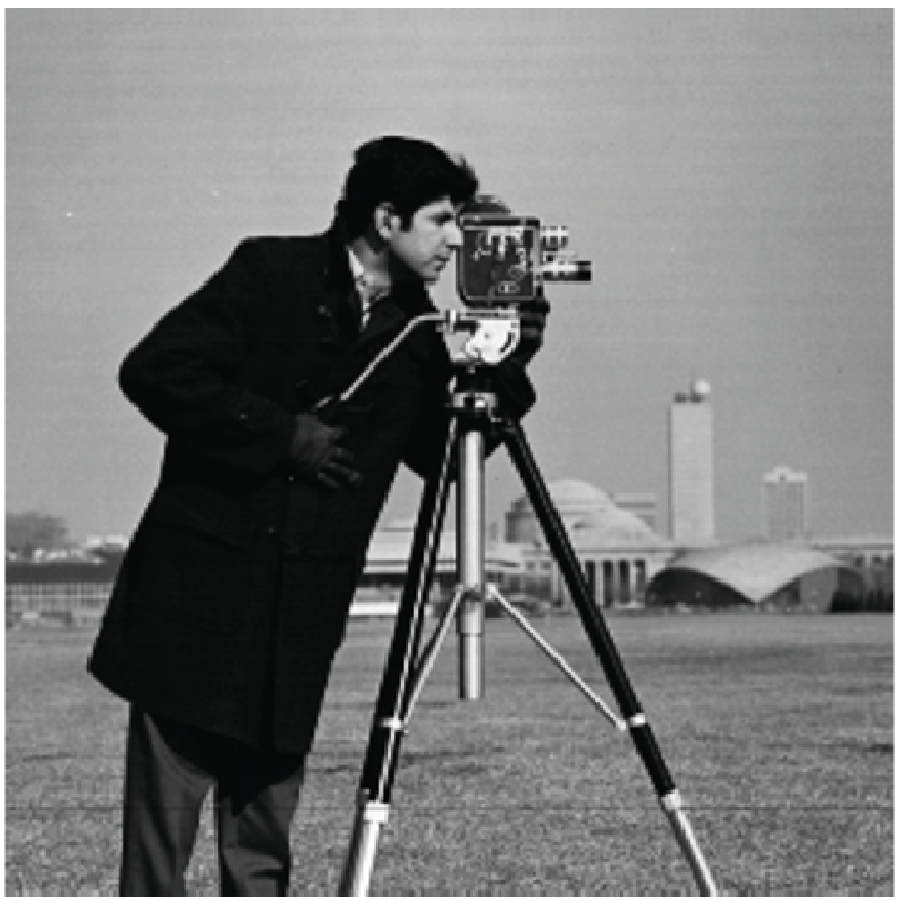}
				
				\label{ori}}
			\subfloat[]{	\includegraphics [width=.3\linewidth]{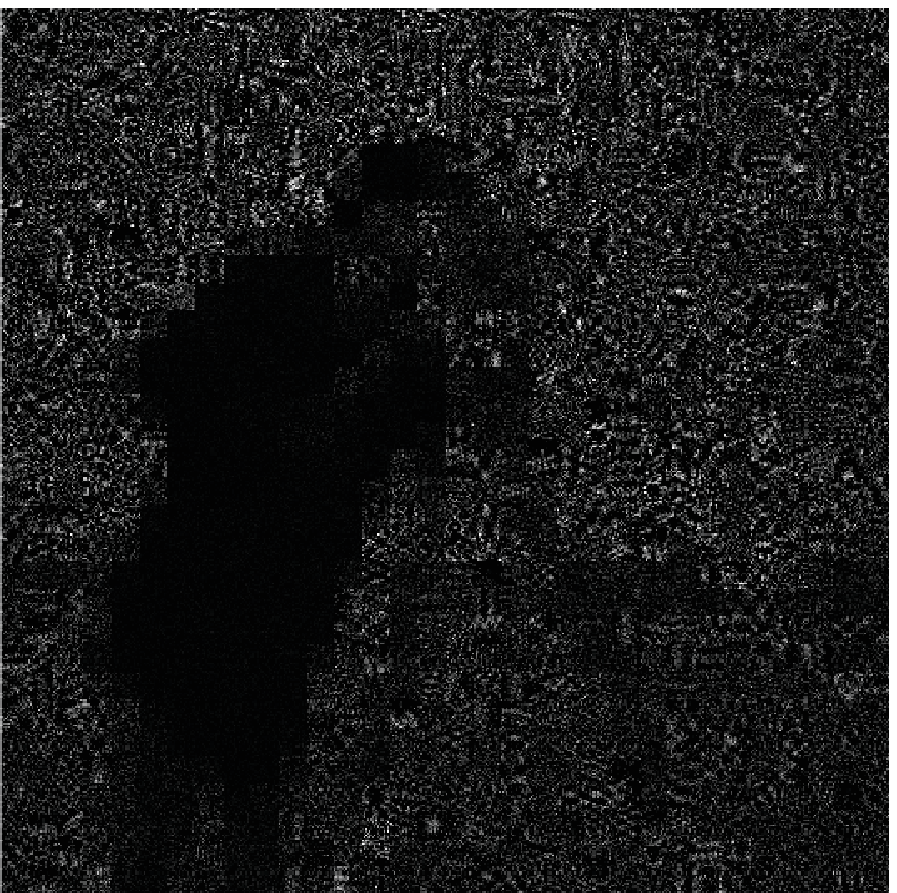}
				
				\label{attack16}}
			\subfloat[]{\includegraphics [width=.3\linewidth]{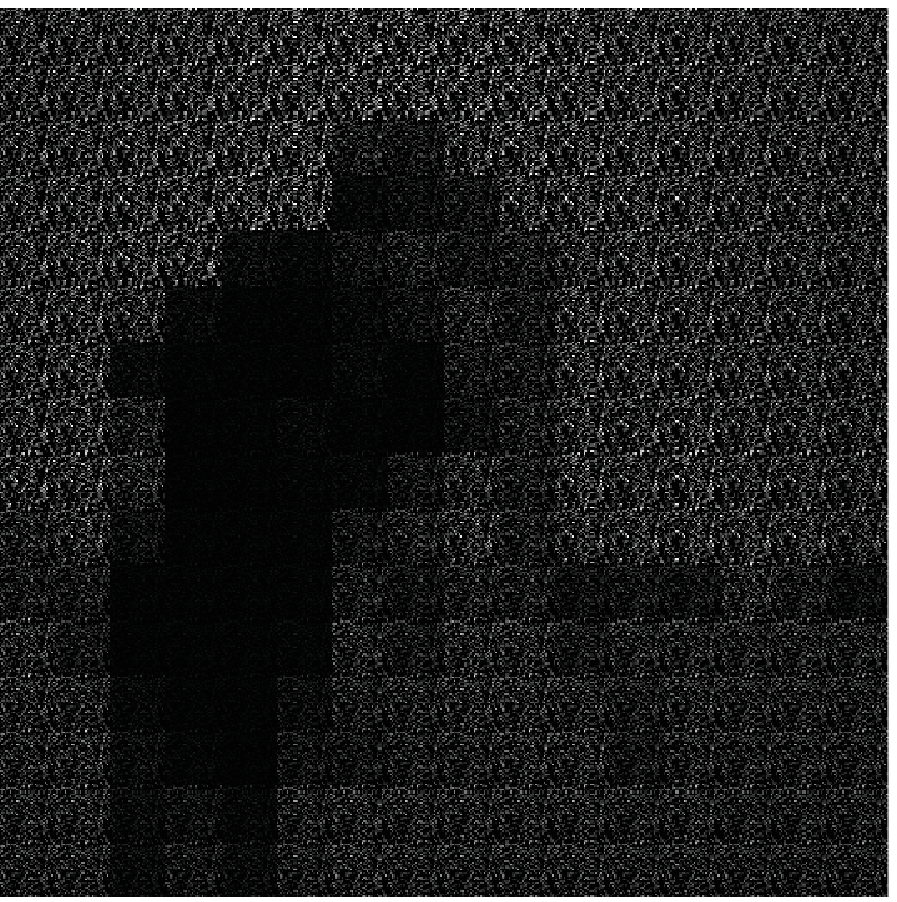}
				
				\label{attack32}}
			\caption{ (a) Original image (b) Ciphertext-only attack on the scheme \cite{SCSDGS},  reconstructed image for  block size $16\times 16$. (c)  Ciphertext-only attack on the schemes \cite{multi2,multi3}, reconstructed image for image block size $32\times 32$.}
		\end{figure}

		The encryption scheme from  \cite{SCSDGS} can be summarized as $\mathbf{y=APx}$, where $\mathbf{A}$ is  the sensing matrix and $\mathbf{P}$ is a permutation matrix and it changes for each sensing. Set $\mathcal{I}$ is a $2K$ largest entries of $abs(\mathbf{A}^T\mathbf{y})$. Using the index set $\mathcal{I}$ a diagonal matrix is generated as:
		\begin{equation*}
		\mathbf{D}_{j,j}=
		\begin{cases}
		1  \text{  if }j\in \mathcal{I} \\
		0   \text{  elsewhere.}
		\end{cases}
		\end{equation*}
		$\mathbf{y}'=\mathbf{y-ADu}'$, here $\mathbf{u'}$ is generated using the secret key (from uniform distribution) and is sent to the cloud. We applied the ciphertext-only attack\cite{OurP} on this scheme for block size $16\times 16$ for image size $512\times512$ and sampling rate $0.25$. We corrupted all the entries of measurement vector by uniform distribution  $U(-10,10)$. We demonstrate the results using cameraman image which is shown in Fig. \ref{ori}. The reconstructed images from the decoding algorithm mentioned in \cite{OurP} using discrete cosine transform as a sparsifying transform are shown in Fig. \ref{attack16}. From Fig. \ref{attack16} it is clear,  from the ciphertext the attacker can retrieve information about the image by applying the ciphertext-only attack. This vulnerability makes this scheme  \cite{SCSDGS} prone to more sophisticated attacks at cloud because sensing matrix is known to the cloud. 
		
		The encryption schemes in \cite{multi2,multi3} are based on one time sensing (OTS) using Binomial sensing matrix. The OTS encryption scheme can be given as,  $\mathbf{y}_i=\mathbf{\Phi}_i \mathbf{x}_i$, where $\mathbf{x}_i$ and $\mathbf{y}_i$ are plaintext and ciphertext. $\mathbf{\Phi}_i$ is sensing matrix for the $i^{th}$ block of the signal. We apply the ciphertext-only attack on these schemes for image size $512\times512$ with block size $32\times32$ and change the sensing matrix for each block. The reconstructed image is shown in Fig. \ref{attack32}. The semi-authorized users can potentially apply such an attack to increase their access level as they know part of the sensing matrix. 
		
		Now  the MPCC scheme is applied to images to provide multi-class encryption as well as privacy preserving cloud computing. 
		Since most of the natural images are not sparse in the time domain. We use data dependent basis  Karhunen-Loeve transform (KLT) because image or video captured in the IoT devices mostly pointing to the same background\cite{klt}. Here it is also assumed that the KLT bases are only known to the data consumers  not to the cloud.  To further reduce the computation at the IoT device we perform block based processing. We consider image of size $512\times 512$ and the block size of $32\times32$. Each block is vectorized to 1024 length signal and for each vectorized block 307 measurements are taken. We achieve the multi-class encryption for images by obfuscating sensitive parts of the image for the semi-authorized user.
		Encryption is performed by first finding the sensitive blocks in the image. Set $\mathcal{S}$ represents the blocks which contain sensitive information then the encryption is performed in the following way, 
		
		\begin{equation}
		\mathbf{y}_i=
		\begin{cases}
		\mathbf{\Phi (r}_i\odot\mathbf{P}_i \mathbf{\Psi} \mathbf{x}_i), & \text{if}\ i\in \mathcal{S} \\
		\mathbf{\Phi (r}_i\odot \mathbf{\Psi} \mathbf{x}_i), & \text{otherwise}.
		\end{cases}
		\end{equation}
		Semi-authorized user's key is $K_r$. Therefore, the semi-authorized user can recover only the non-sensitive blocks of the image,  whereas the superuser can recover the complete signal with its master key $(K_r,K_p)$.
		
		\begin{figure}[htbp]
			
			\subfloat[]{\includegraphics [width=.3\linewidth]{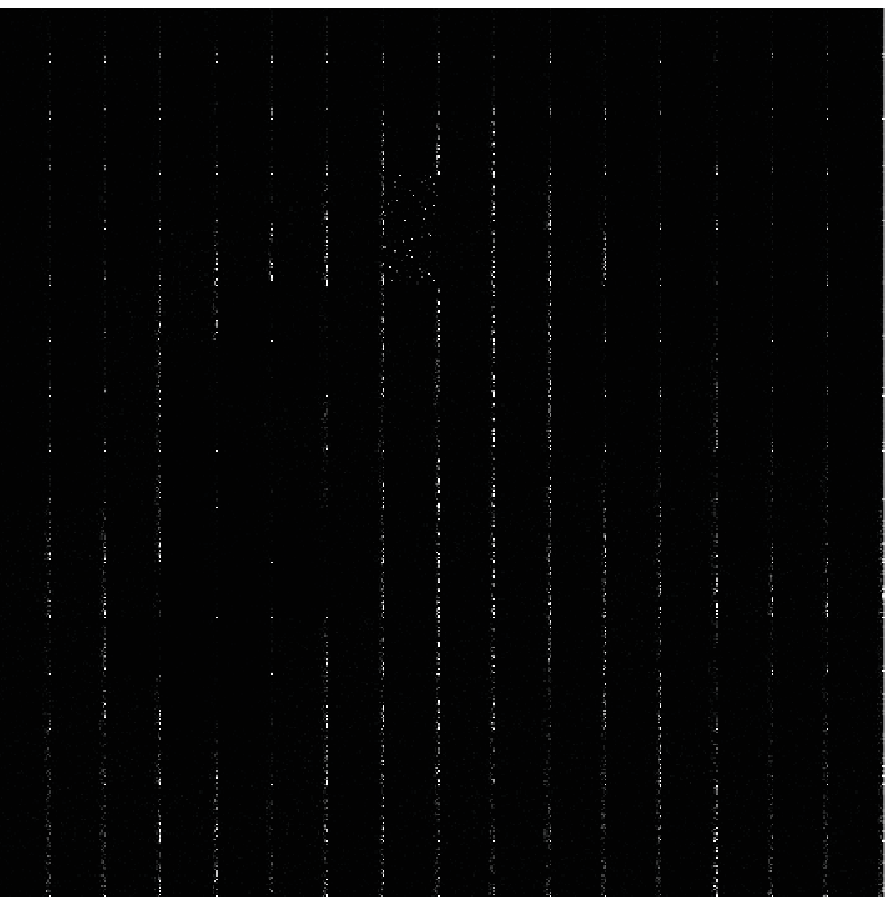}
				\label{cloud}}
			\subfloat[]{	\includegraphics [width=.3\linewidth]{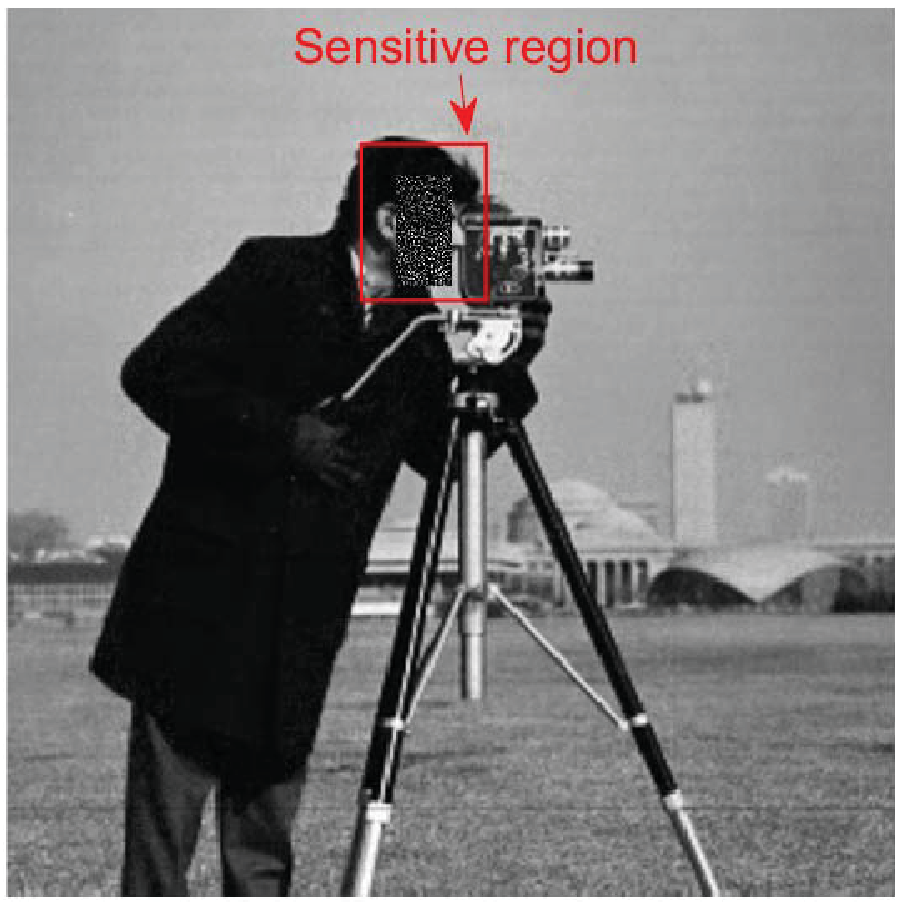}
				
				\label{SAI}}
			\subfloat[]{\includegraphics [width=.297\linewidth]{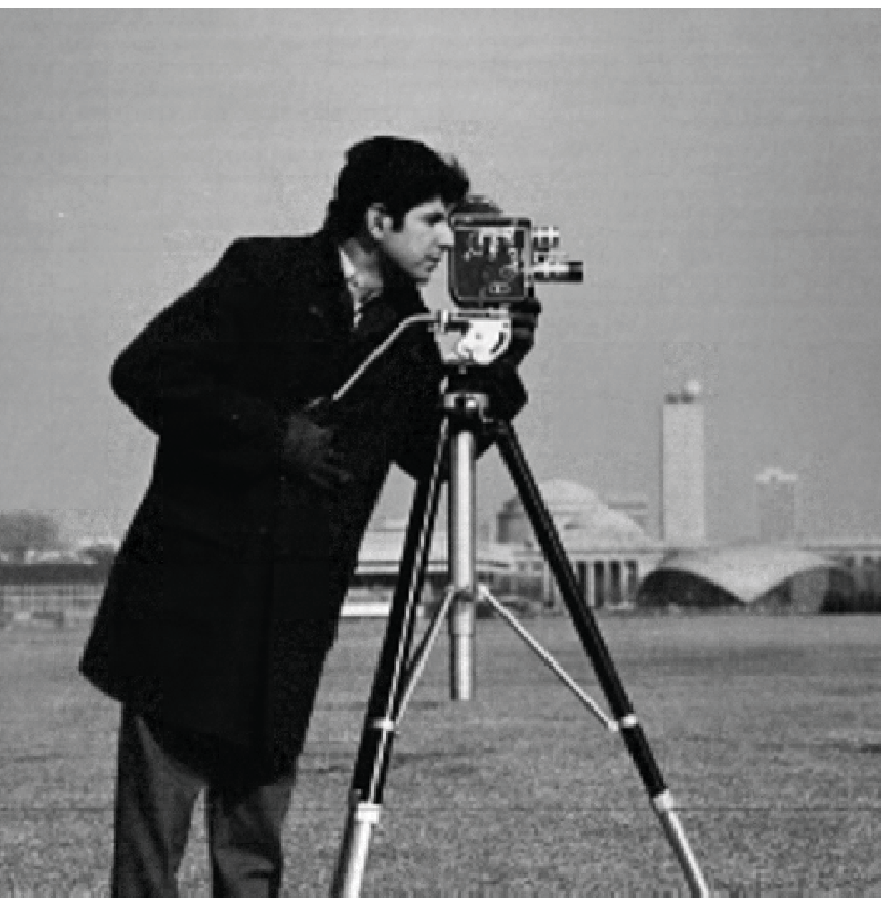}
				\label{SUI}}
			\caption{Application of the MPCC scheme to achieve multi-class encryption. (a) Reconstructed image at the cloud. (b) Reconstructed image for the semi-authorized user. (c) Reconstructed image for the superuser. }
			
		\end{figure}
		%
		%
		For example faces in the image are encrypted using $K_p$ of the master key; therefore, only superuser can retrieve the complete image whereas semi-authorized user can only view image with obfuscated sensitive part. The reconstructed images at cloud, semi-authorized user, and superuser are shown in Fig. \ref{cloud}, Fig. \ref{SAI}, and Fig. \ref{SUI}, respectively. Comparing the reconstructed images at cloud Fig. \ref{cloud} with Fig. \ref{attack16} and Fig. \ref{attack32}, it is clear that the MPCC scheme does not leak any information, which is aligned with the theoretical results.
		\begin{table}[htbp]
			\begin{tabular}{|c|c|c|c|c|}
				\hline
				\multirow{2}{*}{\begin{tabular}[c]{@{}c@{}}Sampling \\ rate\end{tabular}} & \multicolumn{2}{c|}{\begin{tabular}[c]{@{}c@{}}Sensitive  region \\ of the image \\ (PSNR  in dB)\end{tabular}} & \multicolumn{2}{c|}{\begin{tabular}[c]{@{}c@{}}Complete\\  image\\ (PSNR in dB)\end{tabular}} \\ \cline{2-5} 
				& \begin{tabular}[c]{@{}c@{}}Semi-authorized\\ user\end{tabular} & \begin{tabular}[c]{@{}c@{}}Super-\\ user\end{tabular} & \begin{tabular}[c]{@{}c@{}}Semi-authorized\\ user\end{tabular} & \begin{tabular}[c]{@{}c@{}}Super-\\ user\end{tabular} \\ \hline
				0.2 & 0.15 & 0.76 & 24.38 & 29.61 \\ \hline
				0.3 & 0.30 & 2.38 & 25.37 & 36.99 \\ \hline
				0.4 & 0.60 & 6.47 & 25.66 & 47.38 \\ \hline
				0.5 & 1.23 & 14.88 & 25.67 & 55.52 \\ \hline
				0.6 & 2.41 & 32.32 & 25.68 & 62.71 \\ \hline
				0.7 & 4.85 & 67.28 & 25.69 & 66.82 \\ \hline
			\end{tabular}
			\caption{Performance of the reconstructed image for the semi-authorized and superuser using  PSNR  for sensitive and complete image at various
				sampling rates ${M}/{N}$.}
			\label{psnr}
		\end{table}
		
		Reconstruction quality is measured using peak signal-to-noise ratio (PSNR), that is given as,
		$10log\frac{255^2}{\text{MSE}}$,
		where MSE is the mean square error between the original image and the reconstructed image. The PSNR performance for the semi-authorized and the superuser is given in Table \ref{psnr}. From the table, it can be observed that signal quality increases for the superuser as the sampling rate is increased, whereas it remains more or less same for the semi-authorized user. 
		\section{Conclusion}
		In this paper, we design the MPCC scheme that provides multi-class encryption and allows the cloud to perform computationally expensive sparse signal recovery, without compromising data privacy. It is proved that applying
		a ciphertext-only attack on the MPCC scheme is computationally infeasible. We also compare the MPCC scheme with the state-of-the-art schemes and show that MPCC not only has better efficiency in terms of storage, encoding, decoding, and data transmission, but also is secure against ciphertext-only attack.
		\section*{Acknowledgement}
		This work is supported by Innovation Fund Denmark (Grant no.8057-00059B.)  and DIGIT center Aarhus University.

	\end{document}